\documentclass[11pt,letterpaper]{article}
\usepackage[T1]{fontenc}
\usepackage[utf8]{inputenc}
\usepackage{lmodern}
\usepackage{amsmath,amssymb}
\usepackage{graphicx,booktabs,array,caption,fancyvrb,placeins}
\usepackage[margin=0.85in]{geometry}
\usepackage{microtype}
\usepackage{xurl}
\usepackage[hidelinks]{hyperref}
\usepackage{enumitem}
\setlist{nosep,leftmargin=*}
\hypersetup{pdftitle={Deterministic Temporal Reconciliation for Effective-Dated Identity Lifecycle Events},pdfauthor={Pramod Ubbala},pdfsubject={Author preprint with synthetic evaluation}}
\title{Deterministic Temporal Reconciliation for Effective-Dated Identity Lifecycle Events}
\author{Pramod Ubbala\\
\small Independent Researcher, Washington, USA\\
\small \texttt{ubbalap@gmail.com}\\
\small ORCID: \href{https://orcid.org/0009-0005-5816-2998}{0009-0005-5816-2998}}
\date{September 2026}
\begin{document}
\maketitle
\begin{center}\footnotesize\textit{This work has been submitted to the IEEE for possible publication. Copyright may be transferred without notice, after which this version may no longer be accessible.}\end{center}
\begin{abstract}
Identity lifecycle automation processes workforce changes that carry two distinct notions of time: when a business fact becomes effective and when an authoritative source last revised it. Corrections, cancellations, future-dated changes, duplicate delivery, and out-of-order arrival can therefore produce divergent account states when processors rely on transport order or effective time alone. This paper presents a deterministic temporal reconciliation method for effective-dated identity events. The method separates logical business facts from their physical revisions, deduplicates exact replays, selects the authoritative revision of each fact using transaction-time freshness, removes canceled facts, and then applies surviving facts in valid-time order. We establish permutation invariance, replay idempotence, revision dominance, and deterministic snapshot construction under explicit source assumptions. A reproducible synthetic fault-injection benchmark compares the method with arrival-order, effective-time, and transaction-time processing. Across 250,000 generated identity histories and 1,585,589 delivered records, the proposed method reconstructed the separately retained authoritative final state in all tested cases. At 75\% delivery disorder, exact-state accuracy was 42.35\% for arrival-order processing, 85.38\% for effective-time sorting, 76.61\% for transaction-time sorting, and 100\% for the proposed method. A separate 40,000-permutation experiment produced the same final state for every tested permutation under the proposed reconciliation contract.
\end{abstract}

This work received no external funding.

\section{Introduction}
Enterprise identity systems continuously create, modify, disable, re-enable, and remove accounts as people join, move within, leave, and sometimes return to an organization. These operations are security-sensitive. NIST SP 800-53 Rev. 5 explicitly treats account creation, modification, enabling, disabling, removal, and alignment with personnel termination and transfer as account-management concerns [1]. The System for Cross-domain Identity Management (SCIM) standardizes resource schemas and HTTP operations for cross-domain identity provisioning [2], [3]. These standards are necessary for interoperable account management, but they do not prescribe a complete reconciliation rule for mutable, effective-dated upstream business events.

The difficulty is not simply message duplication. A workforce event can describe a future business fact, be corrected after its initial publication, be canceled, and be delivered more than once. The order in which the identity processor receives records can differ from both the order in which the source learned about them and the order in which the business facts become valid. Treating any one of these orders as the only timeline creates ambiguity.

Temporal database research has long distinguished valid time, when a fact is true in the modeled world, from transaction time, when a system stores or learns the fact [4], [5]. Modern stream-processing systems likewise recognize that event time and processing or arrival time are distinct, especially under unbounded and out-of-order delivery [8], [9]. Identity lifecycle automation sits directly at this intersection. A processor must reconstruct a security-relevant state from time-varying facts while tolerating the delivery behavior of a distributed system.

This paper focuses on a narrow question: given a finite set of physical event records for one identity, including revisions, cancellations, and exact replays, what deterministic rule should produce the authoritative identity state at a specified time horizon? We call the resulting contract Deterministic Temporal Reconciliation (DTR). The goal is not to introduce bitemporal databases, idempotency, event-time processing, or logical ordering as new ideas. Those concepts are established. The contribution is a domain-specific reconciliation contract that composes them into an explicit, testable rule for effective-dated identity lifecycle events.

The paper makes four contributions:

1) A formal event model that separates logical business-fact identity, physical event identity, valid/effective time, transaction/modification time, and source sequence.

2) A deterministic canonicalization and application algorithm, DTR, that performs replay deduplication and revision reconciliation before valid-time state transitions.

3) Proof sketches for four correctness properties under explicit assumptions: permutation invariance, replay idempotence, revision dominance, and deterministic snapshot construction.

4) A reproducible synthetic fault-injection benchmark comparing DTR with three intuitive baselines under corrections, cancellations, duplicates, and delivery disorder.

\FloatBarrier
\section{Background And Related Work}
\subsection{Identity Lifecycle And Provisioning}
SCIM defines a standard model and protocol for provisioning and managing identity resources such as users and groups across domains [2], [3]. Its enterprise user schema includes workforce-oriented attributes such as employeeNumber, organization, division, department, and manager [2]. SCIM also defines create, retrieve, replace, patch, delete, bulk, filtering, and related protocol behavior [3]. These mechanisms describe resource interchange and mutation, but effective-dated upstream lifecycle semantics remain an application concern.

NIST account-management guidance highlights why reconciliation errors matter. AC-2 requires organizations to create, enable, modify, disable, and remove accounts according to policy and to align account management with personnel termination and transfer processes [1]. An event processor that mistakenly leaves a terminated identity active is not merely inconsistent. It can preserve access beyond the intended business state. Conversely, an erroneous inactive state can interrupt legitimate access and business operations.

\subsection{Temporal Data}
Snodgrass and Ahn formalized a taxonomy of time in databases, helping distinguish when facts are valid from when they are recorded [4]. Jensen and Snodgrass later provided a detailed semantics of time-varying information and bitemporal relations that capture both valid time and transaction time [5]. SQL:2011 subsequently standardized temporal-table functionality, including application-time and system-versioned concepts [6].

A 2026 systematic review of bitemporal databases found that dual-time modeling remains relevant for data integrity, traceability, and temporal consistency, while also identifying limited industrial-scale empirical validation and implementation gaps [7]. DTR does not propose a new bitemporal storage model. It uses the valid-time/transaction-time distinction as the semantic basis for reconciling mutable lifecycle facts before state application.

\subsection{Ordering In Distributed And Streaming Systems}
Lamport showed that distributed events do not inherently possess a single globally meaningful physical-time order and introduced logical clocks for ordering consistent with causality [10]. Later stream-processing systems such as MillWheel exposed logical time and persistent state while addressing fault tolerance [8]. The Dataflow model explicitly treats unbounded, unordered input, late data, and event-time reasoning as first-class concerns [9]. Earlier work on data streams also formalized application-defined time and the challenges created by disorder, skew, and latency [11], while out-of-order processing architectures showed that systems need not impose input order as their semantic foundation [12]. Babcock et al. framed these timing and processing issues as foundational concerns in data stream systems [13].

These systems establish that arrival order is an unreliable semantic foundation. DTR applies the lesson to identity lifecycle data but adds a separate problem that generic event-time sorting does not solve: multiple physical records can represent revisions of the same logical business fact. Before a valid-time order is meaningful, the system must decide which revision of each logical fact remains authoritative.

Research gap. Existing identity standards define resource-management mechanisms, temporal databases define dual-time semantics, and streaming systems address out-of-order processing. The narrower reconciliation contract studied here is the composition of these ideas for mutable, effective-dated identity lifecycle facts: deduplicate physical replays, resolve revisions using transaction-time freshness, remove canceled logical facts, and then apply the surviving facts in valid-time order.

\FloatBarrier
\section{Problem Model}
For one identity subject, let each physical event record be:

\begin{equation}
e = (u,l,i,k,t_v,t_m,s,r,q,p)
\end{equation}
\begin{table}[!htbp]
\centering\small
\caption{Event model used by DTR}
\renewcommand{\arraystretch}{1.15}
\begin{tabular}{>{\raggedright\arraybackslash}p{\dimexpr0.1\linewidth-2\tabcolsep\relax}>{\raggedright\arraybackslash}p{\dimexpr0.28\linewidth-2\tabcolsep\relax}>{\raggedright\arraybackslash}p{\dimexpr0.62\linewidth-2\tabcolsep\relax}}
\toprule
\textbf{Symbol} & \textbf{Field} & \textbf{Meaning} \\
\midrule
u & Subject key & Identity whose lifecycle is affected \\
l & Logical fact identifier & Stable identifier shared by revisions of one business fact \\
i & Physical event identifier & Unique identifier for one immutable emitted revision; exact replays retain i \\
k & Operation & HIRE, REHIRE, UPDATE, or TERMINATE in the benchmark model \\
t\_v & Valid/effective time & When the business fact takes effect \\
t\_m & Modification/transaction time & When the authoritative source last changed this revision \\
s & Source sequence & Monotonic source ordering value used as a tie-breaker \\
r & Revision number & Revision of logical fact l \\
q & Revision state & ACTIVE or CANCELED \\
p & Payload & Attributes used by the state transition \\
\bottomrule
\end{tabular}
\end{table}

The logical identifier l and the physical identifier i serve different purposes. If a termination is corrected three times, the source has emitted multiple physical records but they represent one mutable business fact. DTR first reconciles those records into one authoritative interpretation of l. By contrast, two distinct terminations or a termination followed by a rehire have different logical identifiers and both may remain on the valid-time timeline.

The model assumes an authoritative source for each logical fact, immutable physical event records once emitted, exact replays preserving the same physical event identifier, and a source sequence that is unique or monotonically increasing within the ordering scope. The sequence is not claimed to be a universal distributed causal clock. It is an authoritative tie-breaker supplied by the source contract.

For the formal claims and benchmark, the source contract is summarized by five assumptions:

\begin{itemize}\item A1. Exact delivery replays preserve the same immutable physical event identifier.\end{itemize}
\begin{itemize}\item A2. All revisions of one mutable business fact preserve the same logical fact identifier.\end{itemize}
\begin{itemize}\item A3. The freshness tuple F(e) is a deterministic total order within each logical fact.\end{itemize}
\begin{itemize}\item A4. The state-transition function is deterministic for the same prior state and canonical event.\end{itemize}
\begin{itemize}\item A5. Snapshot reconstruction is evaluated over the finite event set available for the chosen horizon.\end{itemize}
\subsection{Freshness Order And Application Order}
For revisions of the same logical fact, DTR defines a lexicographic freshness key:

\begin{equation}
F(e) = (t_m,s,r,i)
\end{equation}
The latest revision of logical fact l is the record with maximum F(e). The physical event identifier is included only as a final deterministic tie-breaker for malformed or unexpectedly tied source data.

After revision reconciliation, surviving active facts are applied in valid-time order using:

\begin{equation}
A(e) = (t_v,s,l)
\end{equation}
The separation is essential. Freshness answers, "Which representation of this logical fact is authoritative?" Application order answers, "Where does that authoritative fact belong on the lifecycle timeline?" A newer modification time does not automatically make an event later in valid time.

\subsection{State Transition Model}
The evaluation uses a compact state S = (active, value). HIRE and REHIRE set active=true and can establish a payload value. TERMINATE sets active=false. UPDATE changes the payload only when the identity is active. This deliberately small state machine allows temporal-reconciliation errors to be observed without embedding product-specific provisioning behavior.

For a snapshot horizon h, only facts with t\_v \textless{}= h are eligible for immediate application. Future facts can be persisted as scheduled work by an online implementation. The paper evaluates snapshot reconstruction because it yields a precise reference semantics that can also drive online scheduling.

\FloatBarrier
\section{Deterministic Temporal Reconciliation}
Given a multiset M of physical records for a subject and snapshot horizon h, DTR performs four stages.

\begin{minipage}{\linewidth}
\small\textbf{ALGORITHM 1. Deterministic Temporal Reconciliation snapshot reconstruction}
\begin{Verbatim}[fontsize=\small]
 Input:  multiset M of event records; snapshot horizon h
Output:  deterministic identity state S_h
      1  D <- deduplicate M by physical event identifier i
      2  for each logical fact l in D:
      3      e_l <- arg max F(e) among records with logical_id = l
      4  C <- {e_l | e_l.revision_state = ACTIVE and e_l.t_v <= h}
      5  O <- sort C by A(e) = (t_v, s, l)
      6  S <- initial identity state
      7  for each e in O:
      8      S <- delta(S, e.operation, e.payload)
      9  return S
\end{Verbatim}
\end{minipage}

The ordering of stages matters. Sorting every physical record by effective time before revision reconciliation can place multiple revisions of the same fact adjacent to one another, but it does not determine which revision should semantically exist. An imperative cancellation or correction can therefore depend on which physical revision happens to be processed last among equal-effective-time records.

An online processor can implement the same contract incrementally. The latest revision for each logical fact is indexed by l. A revision with a lower F(e) is stale and cannot replace the current authoritative revision. A newer ACTIVE revision replaces the prior revision. A newer CANCELED revision invalidates that logical fact and removes or neutralizes any scheduled action derived from it. Only the authoritative active facts contribute to the valid-time state projection.

\subsection{Correctness Properties}
Proposition 1. Permutation invariance. For any multiset M and any permutation pi(M), DTR(M, h) = DTR(pi(M), h), assuming immutable records and deterministic total-order keys. Deduplication depends on the set of physical identifiers, revision selection is a max operation within each logical-id group, and final application uses a deterministic sort. None depends on arrival position.

Proposition 2. Replay idempotence. Adding an exact replay of an existing physical event does not change the result. Deduplication by i collapses both copies before revision selection. Therefore DTR(M union \{e\}, h) = DTR(M, h) when e is an exact replay already present in M.

Proposition 3. Revision dominance. Within one logical fact l, only the maximum-freshness revision can influence the canonical timeline. Any stale revision e\_s with F(e\_s) \textless{} F(e\_latest) is excluded before state application. Consequently, late delivery of a superseded correction cannot overwrite the latest authoritative interpretation of that same fact.

Proposition 4. Deterministic snapshot construction. After canonicalization, the application key A(e) defines a total order for all eligible active logical facts under the stated tie-breaker assumptions. Folding a deterministic transition function delta over that sequence produces one state S\_h for the same canonical input set.

\subsection{Complexity}
For n physical records and m surviving active logical facts, hash-based deduplication and grouping require expected O(n) time and O(n) transient space in the batch formulation. Sorting the canonical facts requires O(m log m) time. An incremental implementation can retain only the latest revision per logical fact plus any application-state indexes, reducing the need to hold the entire physical history in the hot path while preserving an append-only audit log separately.

\FloatBarrier
\section{Evaluation}
The benchmark is fully synthetic. It contains no production identities, proprietary schemas, employer data, or production performance measurements. The generator creates a hidden authoritative history first, then emits physical revisions and delivery faults. The hidden history is retained separately from DTR and is used to compute expected final state.

\subsection{Research Questions}
RQ1: How does exact final-state accuracy change as delivery disorder increases?

RQ2: Does the same set of physical records produce different outcomes under different delivery permutations?

RQ3: How often do reconciliation errors produce security-relevant false-active or availability-relevant false-inactive states?

RQ4: Does canonicalization reduce the number of state transitions applied downstream?

\subsection{Synthetic Workload}
Each generated identity begins with a hire. It receives one to four updates with effective times sampled from the lifecycle horizon. A termination is generated with probability 0.55. Conditional on termination, a rehire is generated with probability 0.35 when it fits within the horizon. Corrections are injected with operation-dependent probabilities ranging from 0.12 to 0.35. Generated terminations have a 0.28 probability of cancellation. Corrections preserve the logical fact identifier while receiving new physical event identifiers, source sequence values, modification times, and revision numbers.

Delivery faults are injected after the hidden source history is complete. Exact replays are added with probability 0.10. Delivery disorder probability is varied across 0, 0.10, 0.25, 0.50, and 0.75. When an event is selected for disorder, its delivery rank receives bounded random jitter. Five random seeds are evaluated per disorder level, with 10,000 identities per seed. This yields 250,000 identity histories and 1,585,589 delivered physical records in the primary experiment.

\subsection{Compared Methods}
Four reconciliation strategies were evaluated. The first three isolate common single-ordering choices; DTR separates revision freshness from lifecycle application order.

\begin{table}[!htbp]
\centering\small
\caption{Reconciliation strategies evaluated}
\renewcommand{\arraystretch}{1.15}
\begin{tabular}{>{\raggedright\arraybackslash}p{\dimexpr0.17\linewidth-2\tabcolsep\relax}>{\raggedright\arraybackslash}p{\dimexpr0.3\linewidth-2\tabcolsep\relax}>{\raggedright\arraybackslash}p{\dimexpr0.53\linewidth-2\tabcolsep\relax}}
\toprule
\textbf{Method} & \textbf{Ordering rule} & \textbf{Revision/cancellation handling} \\
\midrule
Arrival & Physical delivery order & Processes every physical revision; cancellations use an imperative inverse. \\
Effective & Effective time, then arrival rank & Processes every physical revision; cancellations use an imperative inverse. \\
Transaction & Modification time, then source sequence & Deterministic, but processes every physical revision and orders distinct facts by source-change time. \\
DTR & Reconcile by F(e), then apply by A(e) & Deduplicates replays, selects latest revision per logical fact, and removes canceled facts. \\
\bottomrule
\end{tabular}
\end{table}

\subsection{Metrics}
Exact final-state accuracy: fraction of identities for which the observed pair (active, value) equals the hidden authoritative state.

False-active rate: fraction for which the processor leaves an identity active when the hidden truth is inactive.

False-inactive rate: fraction for which the processor leaves an identity inactive when the hidden truth is active.

Order-sensitive fraction: fraction of identities that produce more than one distinct final state across 20 different delivery permutations of the same source records.

Records applied per identity: number of records that reach the state-transition function after each method's preprocessing.

\subsection{Results}
Table III reports mean exact-state accuracy across five random seeds. DTR matched the hidden authoritative state in all tested cases under the generator assumptions. Effective-time sorting was substantially better than raw arrival order but remained vulnerable because it did not reconcile revisions before application. Transaction-time sorting was permutation invariant, yet it remained semantically incorrect because modification order does not determine where distinct business facts belong on the valid-time lifecycle.

\begin{table}[!htbp]
\centering\small
\caption{Mean exact final-state accuracy across five seeds, 10,000 identities per seed}
\renewcommand{\arraystretch}{1.15}
\begin{tabular}{>{\raggedright\arraybackslash}p{\dimexpr0.2\linewidth-2\tabcolsep\relax}>{\raggedright\arraybackslash}p{\dimexpr0.2\linewidth-2\tabcolsep\relax}>{\raggedright\arraybackslash}p{\dimexpr0.2\linewidth-2\tabcolsep\relax}>{\raggedright\arraybackslash}p{\dimexpr0.2\linewidth-2\tabcolsep\relax}>{\raggedright\arraybackslash}p{\dimexpr0.2\linewidth-2\tabcolsep\relax}}
\toprule
\textbf{Delivery disorder} & \textbf{Arrival} & \textbf{Effective} & \textbf{Transaction} & \textbf{DTR} \\
\midrule
0\% & 74.684\% & 98.666\% & 76.606\% & 100.000\% \\
10\% & 67.718\% & 95.928\% & 76.860\% & 100.000\% \\
25\% & 58.454\% & 92.500\% & 76.780\% & 100.000\% \\
50\% & 47.770\% & 87.950\% & 76.534\% & 100.000\% \\
75\% & 42.350\% & 85.378\% & 76.608\% & 100.000\% \\
\bottomrule
\end{tabular}
\end{table}

\begin{figure}[!htbp]
\centering
\includegraphics[width=0.85\linewidth,height=0.29\textheight,keepaspectratio]{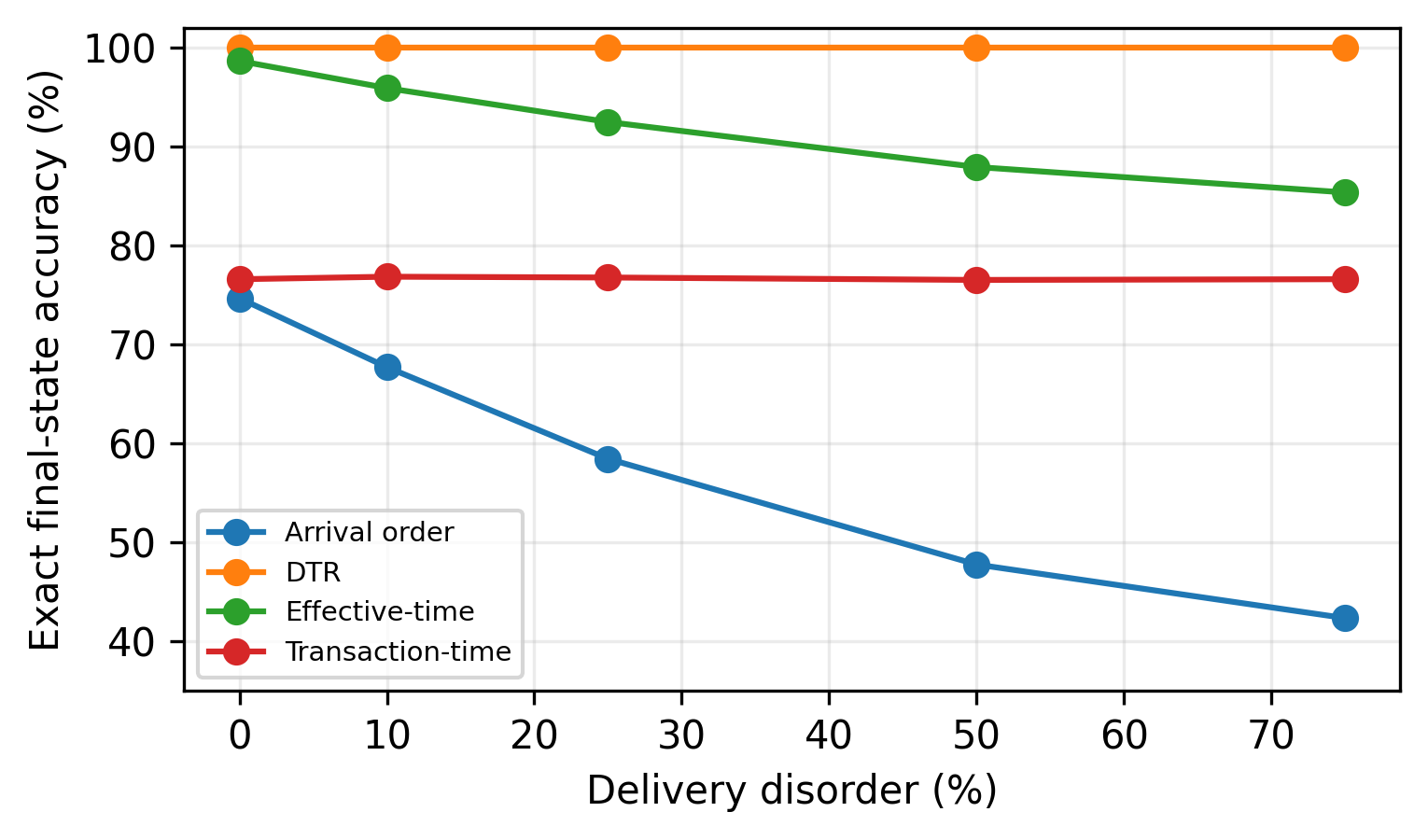}
\caption{Exact final-state accuracy as delivery disorder increases.}
\end{figure}

At 75\% delivery disorder, arrival-order processing left 3.814\% of identities falsely active and 8.774\% falsely inactive. Effective-time sorting produced no false-active cases in this workload but left 3.614\% falsely inactive. Transaction-time sorting left 0.060\% falsely active and 1.120\% falsely inactive. DTR produced neither category of state error in the tested workload.

\begin{figure}[!htbp]
\centering
\includegraphics[width=0.85\linewidth,height=0.29\textheight,keepaspectratio]{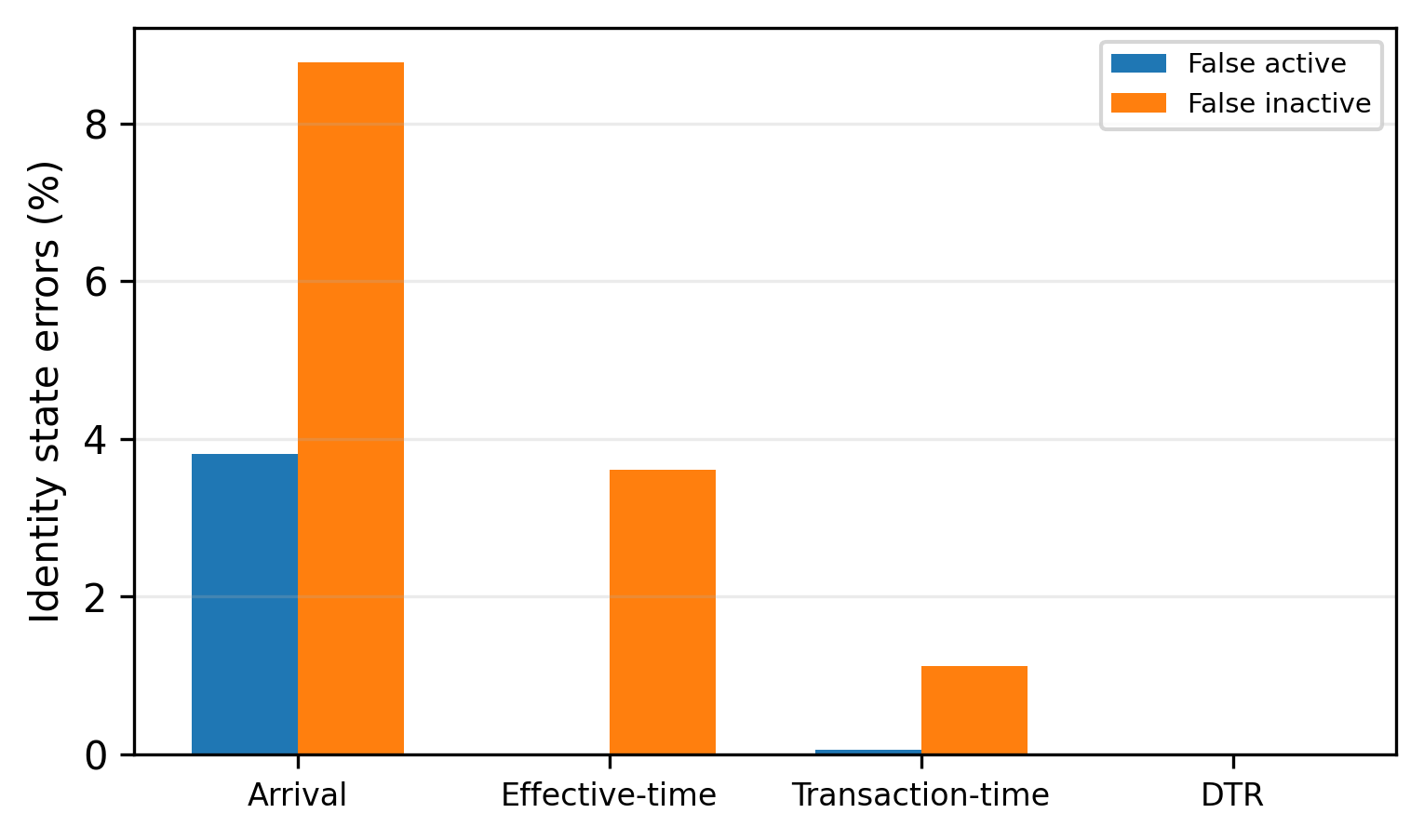}
\caption{Security-relevant state errors at 75\% delivery disorder.}
\end{figure}

A separate order-sensitivity experiment generated 2,000 source histories and delivered each history 20 times under independent full-disorder perturbations, producing 40,000 permutations. The same source records yielded multiple outcomes for 99.0\% of identities under arrival-order processing and 38.7\% under effective-time sorting. Transaction-time sorting and DTR were both permutation invariant, but only 75.7\% of identities were correct under transaction-time sorting across all 20 permutations, compared with 100\% under DTR. This distinction is important: permutation invariance alone does not establish semantic correctness.

\begin{table}[!htbp]
\centering\small
\caption{Sensitivity and correctness across 20 delivery permutations per identity}
\renewcommand{\arraystretch}{1.15}
\begin{tabular}{>{\raggedright\arraybackslash}p{\dimexpr0.18\linewidth-2\tabcolsep\relax}>{\raggedright\arraybackslash}p{\dimexpr0.29\linewidth-2\tabcolsep\relax}>{\raggedright\arraybackslash}p{\dimexpr0.26\linewidth-2\tabcolsep\relax}>{\raggedright\arraybackslash}p{\dimexpr0.27\linewidth-2\tabcolsep\relax}}
\toprule
\textbf{Method} & \textbf{Order-sensitive identities} & \textbf{Mean distinct outcomes} & \textbf{All 20 permutations correct} \\
\midrule
Arrival & 99.0\% & 4.7965 & 1.0\% \\
Effective & 38.7\% & 1.4525 & 61.3\% \\
Transaction & 0.0\% & 1.0000 & 75.7\% \\
DTR & 0.0\% & 1.0000 & 100.0\% \\
\bottomrule
\end{tabular}
\end{table}

\begin{figure}[!htbp]
\centering
\includegraphics[width=0.85\linewidth,height=0.29\textheight,keepaspectratio]{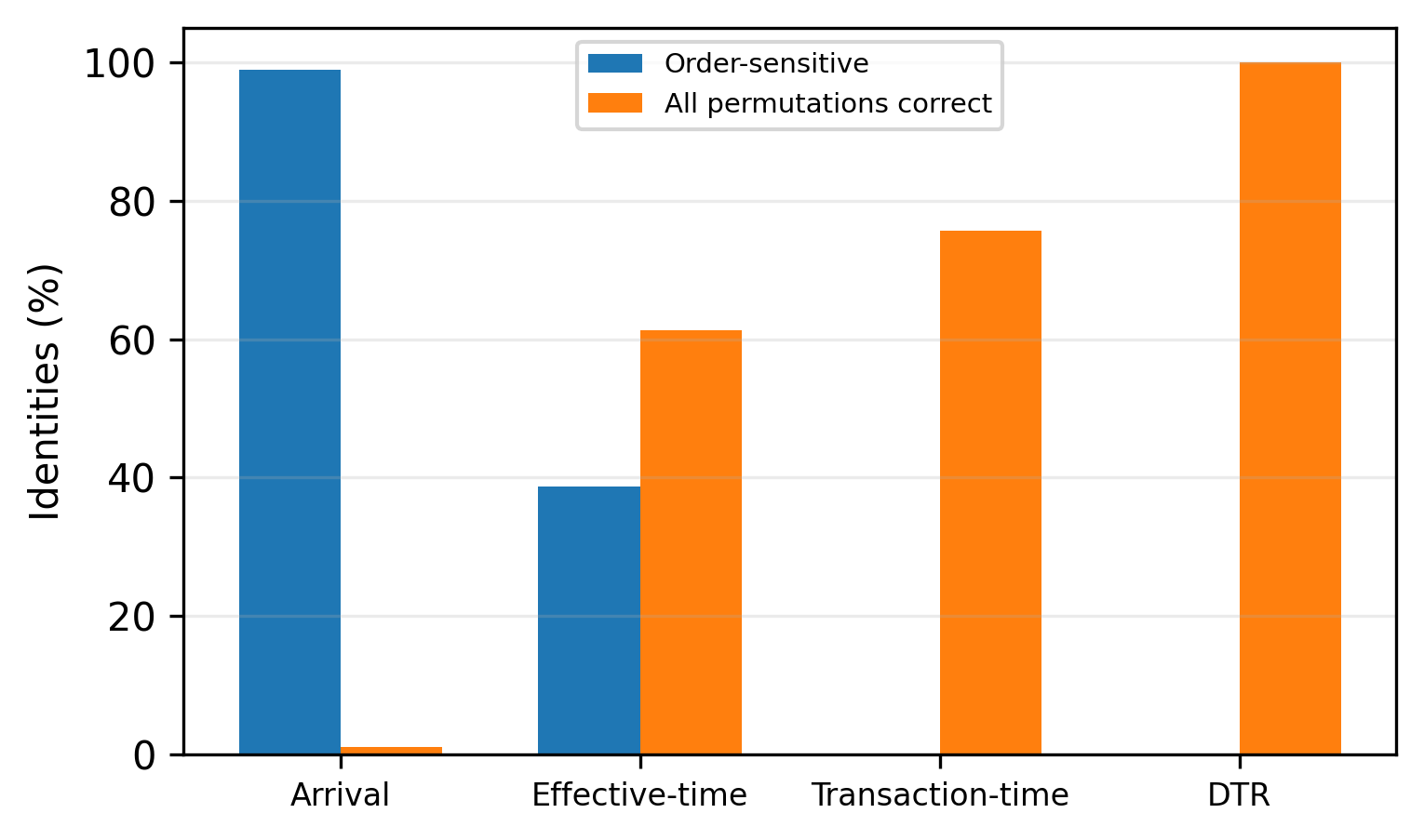}
\caption{Order sensitivity and all-permutation correctness across delivery permutations.}
\end{figure}

Canonicalization also reduced downstream state-transition work. Across the primary experiment, the physical stream contained approximately 6.34 delivered records per identity, while the canonical active timeline contained approximately 4.28 records per identity. DTR therefore avoided about 32.6\% of state-transition applications by removing exact replays, superseded revisions, and canceled logical facts before folding the lifecycle state.

\subsection{Interpretation}
The perfect DTR accuracy observed here should be interpreted narrowly. The generator and DTR share the same declared source contract: the latest transaction-time revision defines the authoritative interpretation of one logical fact, while effective time orders distinct surviving facts. The experiment validates that the implementation reconstructs that contract despite delivery faults. It does not prove that every real identity source follows this contract.

The baselines represent common implementation instincts. Arrival-order processing trusts the transport. Effective-time processing recognizes business time but treats every physical revision as an independent state transition. Transaction-time processing is deterministic but applies distinct business facts in source modification order rather than lifecycle order. The results isolate two separate failure modes: revision reconciliation cannot be replaced by valid-time sorting, and lifecycle construction cannot be replaced by transaction-time sorting.

\FloatBarrier
\section{Discussion}
\subsection{Why Two Temporal Decisions Are Required}
A correction can be newer in transaction time while referring to an earlier effective date. A future termination can be older in transaction time than a correction to today's job attributes. These records should not be globally sorted by modification time. Transaction-time freshness is meaningful within the revision history of the same logical fact. Valid time is meaningful across the surviving facts that compose the identity timeline.

DTR therefore avoids a misleading "one timestamp wins" rule. It uses transaction-time information locally for revision dominance and valid-time information globally for lifecycle construction. This division mirrors the semantic distinction established in temporal database research while remaining implementable with ordinary event stores and message-driven processors.

\subsection{Implications For Identity Security}
Identity lifecycle errors are asymmetric. A false-active result after an authoritative termination can preserve access longer than intended. A false-inactive result can deny legitimate access and generate support or operational incidents. The benchmark therefore reports these states separately instead of relying only on aggregate accuracy.

DTR does not replace downstream authorization controls, session revocation, access reviews, or policy enforcement. It addresses an earlier correctness boundary: whether the provisioning processor has reconstructed the source-authoritative lifecycle state consistently from mutable temporal events.

\subsection{Implementation Guidance}
A practical implementation should make the reconciliation contract explicit rather than implicit in queue behavior. The following patterns follow directly from the model:

\begin{itemize}\item Persist both a stable logical fact identifier and an immutable physical event identifier. One identifier cannot safely serve both revision grouping and replay deduplication.\end{itemize}
\begin{itemize}\item Persist effective time separately from source modification time. Do not infer one from transport arrival time.\end{itemize}
\begin{itemize}\item Treat source sequence/version as part of the authoritative contract, not as an incidental logging field.\end{itemize}
\begin{itemize}\item Canonicalize before issuing side effects. A stale revision should be rejected before account mutation, not compensated after an unnecessary mutation when avoidable.\end{itemize}
\begin{itemize}\item Maintain scheduled future work by logical fact identifier so that a correction or cancellation can replace or remove the scheduled action deterministically.\end{itemize}
\begin{itemize}\item Retain an append-only audit history even if the hot reconciliation index keeps only the latest revision. Operational compactness and forensic traceability are different requirements.\end{itemize}
\subsection{Relationship To Exactly-Once Processing}
DTR does not require exactly-once message delivery. Exact replays are expected and removed by event identifier. Nor does DTR claim that deduplication alone provides exactly-once side effects across arbitrary external systems. Instead, it defines a deterministic desired-state projection. A production processor can combine that projection with idempotent side-effect APIs, compare-and-set state, transactional outbox patterns, or compensating controls appropriate to its environment.

\FloatBarrier
\section{Threats To Validity And Limitations}
Synthetic workload. All evaluation data are generated. The probabilities of corrections, cancellations, termination, rehire, duplicates, and delivery disorder are scenario parameters, not measured population estimates from a real organization.

Contract-level validation. DTR is evaluated against a hidden truth generated from the same declared semantic contract. This is appropriate for testing implementation correctness but does not establish that the contract matches every human-resources system or identity authority.

Single-authority assumption. The model assumes an authoritative ordering scope for each logical fact. Concurrent updates from multiple independent authorities, conflicting writers, or sources without a trustworthy sequence require additional conflict-resolution semantics.

Snapshot-focused metric. The benchmark evaluates final state at a horizon. It does not measure the duration of transient incorrect states, downstream propagation delay, authorization-session revocation latency, or user-visible outage duration.

No production-scale performance claim. The benchmark measures reconciliation correctness and state-transition counts in a Python reference implementation. It does not report throughput, latency, storage cost, cloud cost, or availability of a production service.

Compact state machine. The transition model uses HIRE, REHIRE, UPDATE, and TERMINATE with a simple payload. Real systems can add leave, transfer, conversion, multiple account types, role changes, and policy-dependent transitions. Those operations should be added without changing the separation between revision freshness and valid-time application order.

Cancellation representation. The benchmark represents cancellation as a newer revision of the same logical fact with state CANCELED. Sources that emit compensating events as separate logical facts need an explicit linkage rule before applying DTR.

\FloatBarrier
\section{Conclusion}
Effective-dated identity lifecycle automation is a temporal reconciliation problem as much as a provisioning problem. Delivery order, source modification order, and business effective order encode different semantics. Processing them as one interchangeable timeline makes corrections and cancellations vulnerable to duplicates and late delivery.

DTR makes the reconciliation rule explicit. Physical replays are deduplicated, revisions of the same logical fact are resolved by transaction-time freshness, canceled facts are removed, and surviving facts are applied in deterministic valid-time order. Under the stated synthetic source contract, the reference implementation was invariant to delivery permutation and reconstructed the hidden authoritative state across all tested disorder levels. Effective-time sorting materially improved on raw arrival order but remained order-sensitive because it lacked revision canonicalization. Transaction-time sorting was order-insensitive yet still semantically wrong for a substantial fraction of histories because it used the wrong timeline across distinct business facts.

The next research step is not to add more descriptive lifecycle cases. It is to evaluate the same contract against richer multi-source conflicts, intermediate-state exposure, and implementation-level throughput and latency. That would test whether the semantic benefits demonstrated here remain practical under realistic distributed provisioning workloads.

\FloatBarrier
\section{Reproducibility And Data Availability}
The supplementary artifact accompanying this manuscript contains the complete synthetic benchmark script, raw CSV results, aggregated summary data, and generated figures. No production data, proprietary schemas, employer-identifying information, source code from an operational system, screenshots, internal URLs, or confidential metrics are included.

\section*{Acknowledgment}
The author used OpenAI ChatGPT (GPT-5.6 Sol) to assist with drafting and refinement of Sections I-IX, formalization of the reconciliation algorithm, implementation and checking of the synthetic benchmark, and manuscript formatting. The benchmark results reported in this article were generated from the accompanying reproducibility artifact, and the cited technical sources were checked against publisher, standards-body, or bibliographic records. The author retains responsibility for the final submitted content [14]. No external funding is claimed, and no competing interests relevant to the synthetic evaluation are declared.

\FloatBarrier

\section*{Author}
PRAMOD UBBALA received the B.Tech. degree in information technology from the Indian Institute of Information Technology Allahabad, India, in 2007. He conducts independent research on identity lifecycle automation, distributed systems, cloud security, temporal event processing, and dependable software engineering. His research interests include deterministic provisioning, secure source-of-authority migration, event-driven reconciliation, and correctness in distributed identity workflows.
\end{document}